\documentstyle[psfig]{mn}

\footnotesize
\newdimen\digitwidth    
\setbox0=\hbox{\rm0}
\digitwidth=\wd0
\catcode`!=\active
\def!{\kern\digitwidth}
\normalsize

\newcommand\nuddd{\ifmmode\stackrel{\bf \,...}{\textstyle \nu}\else$\stackrel{\,...}{\textstyle \nu}$\fi}
\def\lsim{~\rlap{$<$}{\lower 1.0ex\hbox{$\sim$}}}

\title{Unusual glitch behaviours of two young pulsars}

\author[Zou, Wang, Wang et al.]{W. Z. Zou,$^{1,2}$\thanks{Email: zouwz@ms.xjb.ac.cn}
~~N. Wang,$^{1,3,4}$  ~~H. X. Wang,$^{1}$   ~~R. N. Manchester,$^{4}$ 
\newauthor X. J. Wu,$^{5}$ ~~J. Zhang$^{1}$ \\
$^{1}$ National Astronomical Observatories, CAS, 40 South
Beijing Road, Urumqi, 830011, China\\
$^{2}$ Graduate School of CAS, 19 Yuquan Road, Beijing, 100039, China\\
$^{3}$ School of Physics, University of Sydney, NSW 2006, Australia\\
$^{4}$ Australia Telescope National Facility, CSIRO, PO Box
76, Epping, NSW 1710, Australia\\
$^{5}$ Astronomy Department, School of Physics, Peking University, Beijing, 100871, China\\}

\begin{document}
\maketitle
\pagestyle{plain}

\begin{abstract}
In this paper we report unusual glitches in two young pulsars, PSR
J1825$-$0935 (B1822$-$09) and PSR J1835$-$1106. For PSR J1825$-$0935,
a slow glitch characterised by a temporary decrease in the slowdown rate
occurred between 2000 December 31 to 2001 December 6. This event
resulted in a permanent increase in frequency with fractional size
$\Delta\nu/\nu\sim31.2(2)\times10^{-9}$, however little effect
remained in slowdown rate.  The glitch in PSR J1835$-$1106 occurred
abruptly in November 2001 (MJD 52220$\pm$3) with
$\Delta\nu/\nu\sim14.6(4)\times10^{-9}$ and little or no change in the
slow-down rate. A significant change in $\ddot\nu$ apparently occurred
at the glitch with $\ddot\nu$ having opposite sign for the pre- and
post-glitch data.
\end{abstract}

\section{Introduction}
A pulsar glitch is a phenomenon in which there is an abrupt increase
in the rotation frequency $\nu$, often accompanied by an increase in
slow-down rate. Typically the fractional increase in pulsar rotation
frequency is in the range $\Delta\nu/\nu=10^{-9}\sim10^{-6}$, and the
relative increment in slow-down rate is
$\Delta\dot\nu/\dot\nu\sim10^{-3}$, where $\nu$ and $\dot\nu$ are
pulsar rotation frequency and frequency derivative,
respectively. Larger glitches in younger pulsars are usually followed
by an exponential recovery or relaxation back toward the pre-glitch
frequency, while for older pulsars and small glitches the jump tends
to be permanent. Glitch activity, defined as the accumulated pulse
frequency change due to glitches divided by the data span,
is high in young pulsars with characteristic age of $10^4$ --
$10^5$~yr, while it is low for very young and very old pulsars
\cite{sl96,wmp+00}.


The trigger of the pulsar glitch is not well understood. In the
classic starquake model, as a consequence of long-term slow-down in
spin rate, deformation stress in the rigid crust builds up to resist
the decreasing oblateness \cite{bppr69}. When the stress exceeds a
critical point, the crust cracks suddenly, resulting in a sudden
increase in spin rate. In the superfluid vortex unpinning
and re-pinning model, triggering of the glitch is due to coupling of
crust and the superfluid interior as a consequence of a sudden
unpinning of vortex lines and the post-glitch relaxation is due to the
vortex gradually re-pinning to the crust lattice
\cite{ai75,acp89}. Based on the observed typical glitches, both of the
models have a sudden increase in rotation frequency and slow-down rate
(i.e.  $\Delta\dot\nu/\dot\nu>0$) at the time of the glitch. The
post-glitch relaxation represents a return to equilibrium with a
linear response of the interior superfluid, while the lack of
relaxation represents a nonlinear response of the superfluid
\cite{acp89,rzc98}.  As more glitches were detected, it became clear
that glitch behaviour varies in aspects such as glitch rate, amplitude
and relaxation. In some cases discrete timing behaviours such as slow
glitches were observed, i.e., the pulsar is spun-up over a time scale
of days, weeks or even months \cite{sha98,wmp+00,wbl01}, accompanied
by decreased slow-down rate ($\Delta\dot\nu/\dot\nu<0$). These diverse
features suggest glitches are triggered locally in the superfluid
interior.

In addition to glitches, pulsars also suffer another kind of timing
irregularity known as timing noise, which is characterised by
restless, unpredictable, smaller scale fluctuations in spin rate
\cite{cd85} with timescales from days to years. The timing noise
induced fluctuations of pulse frequency are small, with fractional
changes $\delta\nu/\nu<10^{-9}$.  Younger pulsars generally show more
timing irregularities \cite{cd85,dsb+98}.

In this paper we report two unusual glitch events in PSR J1825$-$0935
(B1822$-$09) and PSR J1835$-$1106 detected by Urumqi 25~m radio
telescope.  PSR J1825$-$0935 is well known by its rare properties of
interpulse, mode-changing, drifting sub-pulse and microstructure in
the pulse emission (Fowler et al. 1981; Gil et
al. 1994). \nocite{fwm81,gjk+94}  Earlier observations have revealed
unusual glitch events for PSR J1825$-$0935 (Shabanova 1998, Shabanova
\& Urama 2000), in which there was a gradual spin-up or equivalently a
decreased slow-down rate over a few hundred days. In this paper we
present a new but similar discrete timing event which occurred between 2000
December 31 and 2001 December 6. We interpret this as a slow glitch
as suggested by Greenstein (1979) and Cordes
(1979). \nocite{gre79,cor79}

PSR J1835$-$1106 is a young pulsar (characteristic age $\sim 10^5$ yr)
discovered in Parkes Southern Pulsar Survey \cite{mld+96} which has no
previously detected glitch.  In Section~\ref{sec:obs} we introduce the
observations and data analysis, Section~\ref{sec:res} describes the
glitches in detail, and in Section~\ref{sec:dis} we discuss our
results.

\nocite{sha98,su00}

\section{Observations and data analysis}\label{sec:obs}
Observations for PSR J1825$-$0935 and PSR J1835$-$1106 were made
regularly at 1540~MHz as part of the Urumqi Observatory timing program
which commenced in 1999. Currently about 200 pulsars are monitored
with an average interval between observations of about nine days
\cite{wmz+01}. The data for this work consist of observations spanning
$\sim$1400 days from January 2000 to November 2003. The receiver was a
dual-polarisation room-temperature system before July 2000 and was
then upgraded to a cryogenic system. The dedispersion is provided by a
$2\times128\times2.5$ MHz filterbank. The 1-bit digitised data were
sampled at 1-ms intervals and on-line folded at the topocentric period
to form sub-integrations. The offline data reduction includes
dedispersing and summing the data in frequency and time to form mean
pulse profiles.

The mean pulse profile from each observation was cross-correlated
with a high quality template to produce pulse topocentric times of
arrival (TOAs), which were then analysed using the timing program
TEMPO\footnote{see
http://www.atnf.csiro.au/research/pulsar/tempo/}. The JPL ephemeris
DE200 \cite{sta82} was used to correct TOAs to the Solar-System
barycentre. The time-corrected data for a given pulsar were fitted
with the standard spin-down model in which the predicted pulse phase
$\Phi(t)$ is expressed as:
\begin{equation}
\Phi(t) = \Phi_0 + \nu_0(t-t_0) + \frac{1}{2} \dot\nu{_0}(t-t_0)^2 + \frac{1}{6} \ddot\nu_{0}(t-t_0)^3,
\label{eq:spin-down}
\end{equation}
where $\Phi_0$ is the pulse phase at time $t_0$, and $\nu_0$,
$\dot\nu_0$ and $\ddot\nu_0$ are the pulsar rotation frequency,
frequency derivative and frequency second derivative at time $t_0$
respectively.

In a classic glitch, there is a sudden deviation in the observed pulse
phase at the time of the glitch due to the jump in frequency and
frequency derivative, plus an exponential decay in the frequency jump
as a function of time. However as will be described in more detail in
Section~\ref{sec:1825}, the glitch in PSR B1825$-$0935 builds up
gradually and we describe the glitch effect by comparing the timing
solutions away from the spin-up event. For PSR J1835$-$1106, the
analysis shows an abrupt jump in frequency but little change in
frequency derivative at the glitch epoch, and no exponential decay.

\begin{table*}
\begin{minipage}{10cm}
\caption{The parameters of PSR J1825$-$0935 and PSR J1835$-$1106.}
\begin{tabular}{cclllc}
\hline
PSR J      & PSR B  & ~RA(J2000)    & ~~DEC(J2000)      &  \multicolumn{1}{c}{DM}   & Age
\footnotetext{$^{a}$ Arzoumanian et al., 1994  ~~$^{b}$ Hobbs et al, 2003  ~~$^{c}$ D'Amico et al., 1998}\\
           &        & ~(h m ~s)      & ~~~(d m ~s)         & (cm$^{-3}$ pc) &  (10$^5$yr)  \\
\hline
1825$-$0935 & 1822$-$09  & 18:25:30.596(6)$^{a}$     & $-$09:35:22.8(4)$^{a}$ &  19.39(4)$^{b}$ &  2.33 \\
1835$-$1106 &            & 18:35:18.287(2)$^{c}$     & $-$11:06:15.1(2)$^{c}$ &  132.679(3)     &  1.28 \\
\hline
\end{tabular}
\label{tb:postn}
\end{minipage}
\end{table*}

\nocite{antt94}
\nocite{dsb+98}
\nocite{hlk+03}

Table~\ref{tb:postn} gives the pulsar names and J2000 positions in the
first three columns, the remaining columns contain the pulsar
dispersion measure (DM) and characteristic age.  Uncertainties in the last quoted
digit are given in parentheses.  Given these parameters and rotation
model in Equations~\ref{eq:spin-down} we obtained
the glitch parameters which will be described in detail in next
section. The positions and DMs were held fixed in the fitting process.

\section{Results}\label{sec:res}
\subsection{PSR J1825$-$0935 (B1822$-$09)}\label{sec:1825}
\begin{figure}
\centerline{\psfig{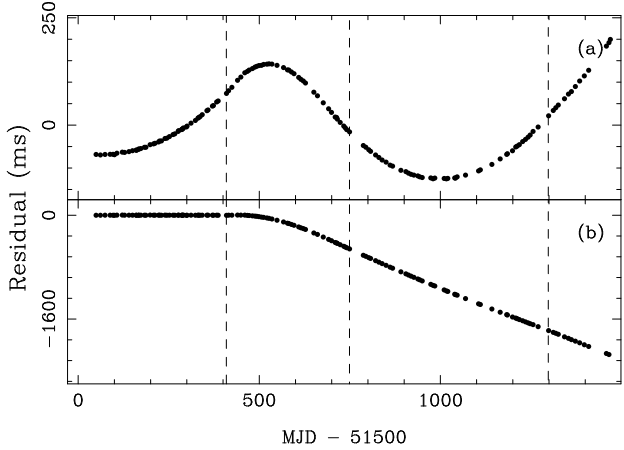}}
\caption{Timing residuals for PSR J1825$-$0935 spanning the whole data
set (a) after fitting for $\nu$ and $\dot\nu$ and (b) with respect to the
model before the slow glitch. The slow glitch occurred between MJD
51909--52249 as indicated by the first two dashed lines. Another
possible slow glitch with smaller amplitude occurred near the end of the data
set as indicated by the third dashed line.}
\label{fg:1825res}
\end{figure}

Fig.~\ref{fg:1825res}(a) shows the timing residuals spanning the whole
data set after fitting for $\nu$ and $\dot\nu$. The three dashed lines,
representing MJDs 51909, 52249 and 52798 respectively, divide the data
into four sections. Large residuals arise when fitting data across the
dashed lines and a change in $\dot\nu$ between the first two dashed lines
is evident. The rotation parameters derived from each section are
given in Table~\ref{tb:rotn}. Fig.~\ref{fg:1825res}(b) shows the
timing residuals for the whole data set with respect to the timing
solution for the first section. The curvature of the residuals
between the first two dashed lines (MJD 51909$-$52249) reveals a gradual
spin-up or, more accurately, a decrease in the spin-down rate for PSR J1825$-$0935.

\begin{figure}
\centerline{\psfig{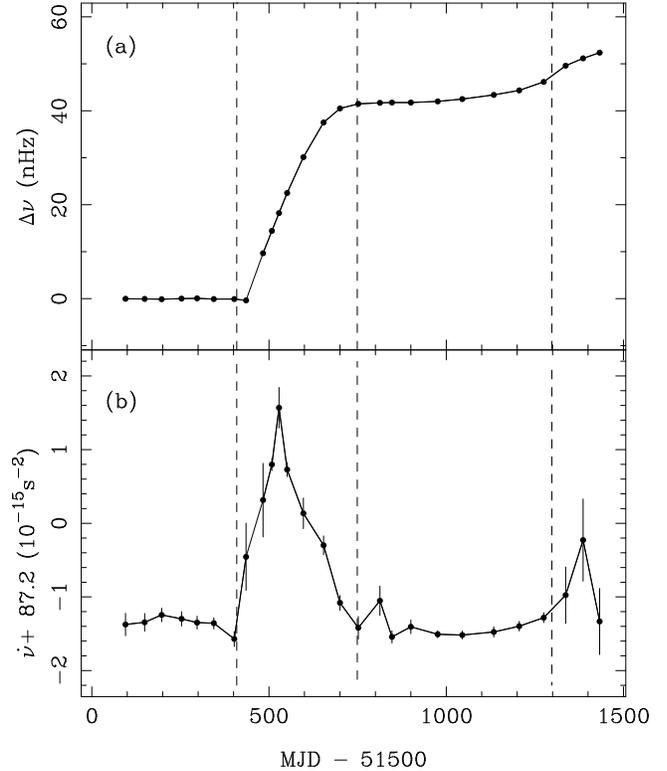}}
\caption{Variation of $\nu$ and $\dot\nu$ for
PSR J1825$-$0935. (a) Frequency residuals $\Delta\nu$ relative to
the pre-glitch solution. (b) Observed variations of $\dot\nu$. }
\label{fg:1825glt}
\end{figure}

This decrease in spin-down rate is clearly shown in
Fig.~\ref{fg:1825glt}, which shows the variations of rotation
frequency and frequency derivative with respect to timing solution
before MJD 51909. The individual values of $\nu$ and $\dot\nu$ in the
plot are derived from short fits spanning 50--100~d. The plot confirms
the gradual increase of pulsar frequency between MJD 51909--52249, and
another possible similar but much smaller event near the end of the
data, as indicated by the third dashed line. Following Greenstein
(1979), we describe these events as slow glitches. As shown in
Fig.\ref{fg:1825glt}(a), for the first slow glitch, there was a
continuous increase in frequency for about 300~d followed by a return
to the initial state for the next 500~d. Associated with the spin-up
process is a decreasing slow-down rate (increasing $\dot\nu$) which
lasted $\sim120$~d with maximum
$\Delta\dot\nu\sim3\times10^{-15}$~s$^{-2}$. The slow-down rate then
decayed to approximately the pre-glitch level within $\sim220$~d.  The
fractional changes in frequency and frequency derivative before and
after the slow glitch are $\Delta\nu/\nu=31.2(2)\times10^{-9}$ and
$\Delta\dot\nu/\dot\nu=1.9(1)\times 10^{-3}$ respectively. These
values and the approximate size of the second possible slow glitch are
given in Table~\ref{tb:glt}.

\begin{table*}
\begin{minipage}{14cm}
\caption{The rotation parameters for PSR J1825$-$0935 and PSR
J1835$-$1106. The errors are at 2$\sigma$ level.}
\begin{tabular}{clllcccc}
\hline
PSR J      & \multicolumn{1}{c}{$\nu$}            & \multicolumn{1}{c}{$\dot\nu$}            & \multicolumn{1}{c}{$\ddot\nu$}           & Epoch   & Fit Span    & Residual  & No. of \\
           &\multicolumn{1}{c}{($s^{-1}$)}        & \multicolumn{1}{c}{($10^{-12}$s$^{-2}$)} & \multicolumn{1}{c}{($10^{-24}$s$^{-3}$)} & (MJD)   & (MJD)       & (ms)     & TOAs\\
\hline
1825$-$0935 & 1.30039948201(3) & $-$0.088487(8)        & --                   & 51718.0 & 51549-51886 & 0.73      & 49\\
           & 1.3003967479(2) & $-$0.08684(2)         & $-$9.0(7)             & 52079.0 & 51909-52249 & 1.46      & 40\\
           & 1.30039331524(3) & $-$0.088657(5)        & --                   & 52529.0 & 52287-52769 & 0.87      & 44\\
           & 1.3003906000(2)  & $-$0.08850(7)         & --                   & 52884.0 & 52798-52969 & 0.93      & 14\\
1835$-$1106 & 6.0273144867(3)  & $-$0.74902(2)         & 96(4)                & 51909.0 & 51600-52218 & 1.34      & 68\\
           & 6.0272702839(4)  & $-$0.74807(2)         & $-$74(4)              & 52595.0 & 52221-53021 & 2.12      & 70\\
\hline
\end{tabular}
\label{tb:rotn}
\end{minipage}
\end{table*}

\subsection{PSR J1835$-$1106}
\begin{figure}
\centerline{\psfig{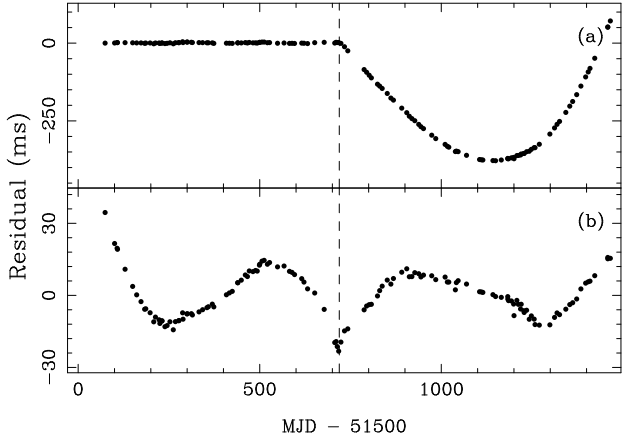}}
\caption{Timing residuals for PSR J1835$-$1106 (a) with respect to the
timing solution before MJD 52220 and (b) after fitting for a glitch at MJD 52220.}
 \label{fg:1835res}
\end{figure}

Fig.~\ref{fg:1835res}(a) shows the pre-fit residuals spanning the
whole data set with respect to the timing solution before MJD
52220. The sudden change in residual slope after MJD 52220 indicates a
clear glitch at this time. The residuals after fitting for a glitch
model with jumps in frequency and frequency derivative at MJD 52220 as
well as the pulse frequency and its first two derivatives are shown in
Fig.~\ref{fg:1835res}(b). Significant cubic terms with opposite signs
are present in the pre- and post-glitch data, suggesting a sign change
in $\ddot\nu$ at the time of the glitch. Fitting separately to the data
sets before and after MJD 52220 gave the solutions in
Table~\ref{tb:rotn} showing this sign change.  

\begin{figure}
\centerline{\psfig{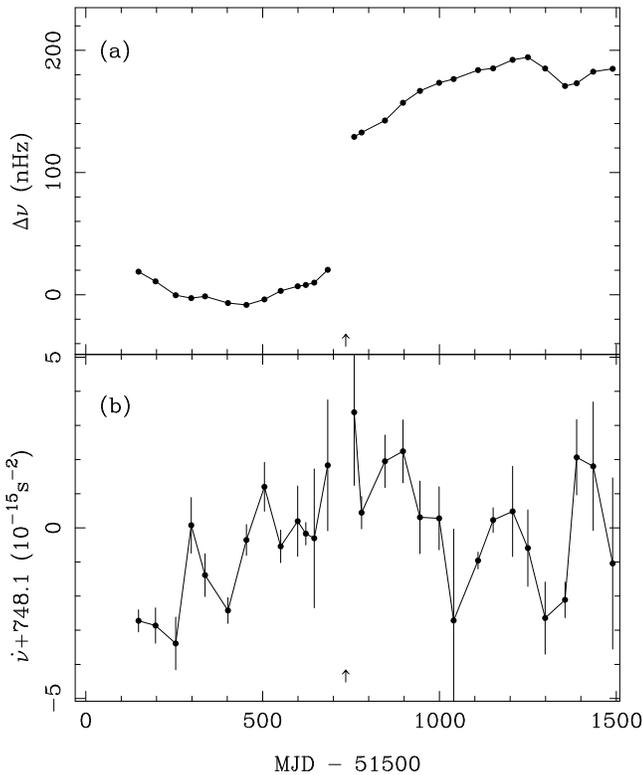}}
\caption{Variation of $\nu$ and $\dot\nu$ across the glitch of PSR
J1835$-$1106. (a) The frequency residual $\Delta\nu$ after subtracting
a fit of $\nu$ and $\dot\nu$ to the pre-glitch data, and (b) Observed
variations of $\dot\nu$.}
\label{fg:1835glt}
\end{figure}

Fig.~\ref{fg:1835glt} presents the variations of $\nu$ and $\dot\nu$
relative to a fit of $\nu$ and $\dot\nu$ to the pre-glitch data.  This
plot confirms the jump in $\nu$ at MJD 52220 and the reversed sign of
$\ddot\nu$ before and after the glitch. There was little or no change
in $\dot\nu$ at the time of the glitch. A coherent timing fit across
the glitch gives the fractional jumps in frequency and frequency
derivative at MJD 52220 listed in Table~\ref{tb:glt}.

\begin{table}
\begin{minipage}{80mm}
\caption{The glitch parameters of PSR J1825$-$0935 and PSR J1835$-$1106.
The errors are at 2$\sigma$ level.}
\begin{tabular}{cllll}
\hline
PSR J      & \multicolumn{1}{c}{Glitch epoch}  & \multicolumn{1}{c}{Date}        & \multicolumn{1}{c}{$\Delta\nu/\nu$}    & \multicolumn{1}{c}{$\Delta\dot{\nu}/\dot{\nu}$} \\
           & \multicolumn{1}{c}{(MJD)}         &             & \multicolumn{1}{c}{($10^{-9}$)}        & \multicolumn{1}{c}{($10^{-3}$)}  \\
\hline
1825$-$0935 & 51909--52249  & 001231--011206   & 31.2(2)      & 1.9(1)\\
            & 52798--52969  & 030608--031126   & $\sim$2.1(5) & $\sim$--1.8(8) \\  
1835$-$1106 & 52220(3)      & 011107      & 14.6(4)       & $-$1.0(2)       \\
\hline
\end{tabular}
\label{tb:glt}
\end{minipage}
\end{table}

\section{Discussion}\label{sec:dis}
Shabanova \& Urama (2000) \nocite{su00} discussed four glitches in PSR
J1825$-$0935 between 1994 September to 1999 February. However we
interpret them as two slow glitches, with the first
occured during MJD 49940 to 50557 (glitches 2a to 2b of Shabanova \&
Urama, 2000), and the second slow glitch beginning at 51054 but still
not completed at the end of their data set (glitch 4 in Shabanova \&
Urama, 2000). Shabanova \& Urama (2000) state that each slow glitch
was preceded by a small glitch (glitches 1 and 3 in their paper) with
fractional size $10^{-10}$, however we did not detect such an event
before the third slow glitch we report in this paper. In the first
slow glitch, the continuous increase in pulsar frequency lasted for
620~d, leaving a permanent increase in frequency with amplitude
16~nHz. The second slow glitch lasted at least 120~d, with frequency
increase $\geq 9$~nHz. The third event reported here lasted $\sim
340$ days and is similar to the previous two but with a much larger
amplitude of 46~nHz.  The possible fourth slow glitch lasted at least
170~d with frequency increase $\geq 3$~nHz. These spin-up events are
separated by 1114~d, 909~d and 549~d respectively. 

It appears that PSR J1825$-$0935 switches between normal steady
slow-down and intervals of decreased slow-down rate. Timing noise in
pulsars is usually attributed to fluctuations in the interior neutron
superfluid and its pinning to the neutron-star crust
\cite{anp86,rud91}. These two phases may represent different states of
the interior superfluid or, less likely, of the magnetospheric
configuration.

The glitch detected in PSR J1835$-$1106 has the normal abrupt increase
in pulsar rotation frequency but little change in $\dot\nu$ at the
glitch epoch, similar to small glitches detected in other
pulsars. A more interesting feature is the apparent reversal of the
sign of $\ddot\nu$ at the time of the glitch. Again, this most likely
originates from a change in the properties of the interior superfluid
at the time of the glitch.

\section{Summary}
In this paper we present two unusual glitches observed recently in PSR
J1825$-$0935 and PSR J1835$-$1106 at Urumqi Astronomical Observatory.
These two glitches are both different to glitches in most other
pulsars, demonstrating the great diversity of glitch behaviours.

The main aspects for PSR J1825$-$0935 glitches are as follows.
They are slow glitches, with the pulsar frequency continuously
increasing for several hundred days, similar to the glitch events
reported by Shabanova \& Urama (2000).  The persistent increases in
frequency result from a temporarily decreased slow-down rate lasting
several hundred days, after which the slow-down rate returns to its
stable value. The main event reported here results in a permanent
frequency increase of 46~nHz, several times larger than the previously
reported events. 

For J1835$-$1106 there was a clear glitch of relative size $\sim
10^{-8}$ at MJD 52220, the first glitch observed in this
pulsar. Unlike glitches observed in most other pulsars, there was no
increase in slow-down rate at the time of the glitch. There was
however a reversal in sign of $\ddot\nu$ apparently associated with
the glitch, indicating a change in the properties of the process(es)
responsible for timing noise in this pulsar.

\section*{ACKNOWLEDGMENTS}
We thank the engineers who maintained the receiver and telescope at
Urumqi Observatory and those who helped with the observations. WZZ
thanks J. Yang and X. F. Wang for assistance. We thank the support
from NNSFC under the project 10173020.


\end{document}